\documentclass[journal]{IEEEtran}

\usepackage[numbers]{natbib}
\usepackage{graphicx}          
\usepackage{amsmath,amssymb,amsfonts}
\usepackage{hyperref}
\hypersetup{hidelinks=true}
\usepackage{textcomp}
\usepackage{algorithm}
\usepackage{algpseudocode}
\newcommand\NoDo{\renewcommand\algorithmicdo{}}
\newcommand\NoThen{\renewcommand\algorithmicthen{}}
\algblock{ParFor}{EndParFor}
\algnewcommand\algorithmicparfor{\textbf{parfor}}
\algnewcommand\algorithmicendparfor{\textbf{end\ parfor}}
\algrenewtext{ParFor}[1]{\algorithmicparfor\ #1}
\algrenewtext{EndParFor}{\algorithmicendparfor}
\usepackage{wrapfig}
\usepackage{xcolor}
\usepackage{subfigure}
\usepackage{mathtools}
\usepackage{amsthm}

\def\tsc#1{\csdef{#1}{\textsc{\lowercase{#1}}\xspace}}
\tsc{WGM}
\tsc{QE}

\newtheorem{theorem}{Theorem}
\newtheorem{lemma}{Lemma}

\newtheorem{definition}{Definition}

\newtheorem{problem}{Problem}
\newtheorem{assumption}{Assumption}
\graphicspath{{Figures/}}
\DeclareMathOperator*{\argmax}{arg\,max}

\begin{document}
\title{Controlled Invariant Sets for Gaussian Process State Space Models}

\author{Paul~Griffioen, Bingzhuo~Zhong, Murat~Arcak, Majid~Zamani, and Marco~Caccamo%
\thanks{P. Griffioen is with the Engineering Department, Dordt University, Sioux Center, IA 51250, USA (e-mail: paul.griffioen@dordt.edu).}%
\thanks{B. Zhong is with the Thrust of Artificial Intelligence, Information Hub, Hong Kong University of Science and Technology (Guangzhou), Guangzhou, Guangdong 511453, China (e-mail: bingzhuo.zhong@hkust-gz.edu.cn).}%
\thanks{M. Arcak is with the Department of Electrical Engineering and Computer Sciences, University of California, Berkeley, CA 94720, USA (e-mail: arcak@berkeley.edu).}%
\thanks{M. Zamani is with the Department of Computer Science, University of Colorado Boulder, Boulder, CO 80309, USA (e-mail: majid.zamani@colorado.edu).}%
\thanks{M. Caccamo is with the School of Engineering and Design, Technical University of Munich, Garching 85748, Germany (e-mail: mcaccamo@tum.de).}%
\thanks{This work was supported by the Air Force Office of Scientific Research under Grant FA9550-21-1-0288 and the National Science Foundation under Grants CNS-2111688, CNS-2145184, and CNS-2039062. M. Caccamo was supported by an Alexander von Humboldt Professorship endowed by the German Federal Ministry of Education and Research.}%
\thanks{P. Griffioen and B. Zhong contributed equally to this work.}%
}





\maketitle

\begin{abstract}
We compute probabilistic controlled invariant sets for nonlinear systems using Gaussian process state space models, which are data-driven models that account for unmodeled and unknown nonlinear dynamics.  We propose a semidefinite programming scheme for designing state-feedback controllers that maximize the probability of the trajectories staying within a probabilistic controlled invariant set while satisfying input constraints.
The results are validated on a quadrotor, both in simulation and on a physical platform.
\end{abstract}

\section{Introduction}
 Gaussian Process State Space Models (GPSSMs) are 
 increasingly used to account for the nonlinearities and unknown dynamics of physical systems \cite{umlauft2020learning,eleftheriadis2017identification,frigola2014variational,turner2010state,frigola2013bayesian,svensson2016computationally}. In contrast to parametric models like recurrent neural networks, GPSSMs are inherently regularized by a prior model, mitigating the tendency to overfit. Furthermore, GPSSMs quantify uncertainty and modeling errors as a distribution over functions, ensuring that the model is not overconfident in regions of the state space where data is scarce \cite{schneider1996exploiting,deisenroth2013gaussian}. GPSSMs are commonly employed in safety critical applications \cite{xie2020learning,yin2022gaussian,zeng2019gaussian}, in which system failures, such as collisions, may result in catastrophic consequences. For these applications, it is crucial to design controllers that prevent the system from reaching unsafe regions of the state space.

To ensure safety, abstraction-based approaches have been studied in \cite{zhong2023automata,Ding2013stochastic,Abate2008Probabilistic,Kamgarpour2011Discrete}. Instead of considering continuous state and input sets, these approaches discretize the state and input sets of the original system to construct finite abstractions. However, the number of discrete states and inputs grows exponentially with the dimensions of the state and input sets,  limiting applicability to high-dimensional systems. In contrast, invariant sets \cite{blanchini1999set} can be computed without an abstraction to ensure the safety of the system under the effects of uncertainties. In particular, robust invariant sets \cite{brockman1995quadratic,brockman1998quadratic,alessandri2004estimation} are used to describe a region of the state space in which the trajectory is guaranteed to remain under  bounded disturbances. Similarly, probabilistic invariant sets \cite{kofman2012probabilistic,kofman2016continuous,pizzi2021probabilistic} have been proposed to describe a set containing the trajectory  at all times with a certain probability. A relationship between robust and probabilistic invariant sets for linear systems is established in \cite{hewing2018correspondence}.

Other results, such as \cite{wang2018safe,taylor2020learning}, provide robust invariance guarantees via barrier functions, but they do not treat the uncertainty as probabilistic. Instead, uncertainty is treated as a fixed function for which the Gaussian Process (GP) regression model provides a prediction with pointwise error bounds. The result in \cite{griffioen2023probabilistic} extends the probabilistic invariance results in \cite{kofman2012probabilistic,kofman2016continuous,hewing2018correspondence} to nonlinear systems in the form of GPSSMs but does not synthesize controllers to maximize the probability of positive invariance. Data-driven control synthesis for uncertain control systems is addressed in \cite{Zhong2022Synthesizing,mulagaleti2021data,danielson2016path}; however, the results only apply to linear systems with bounded disturbances instead of nonlinear systems containing stochastic uncertainty.

In this paper, we propose a semidefinite programming (SDP) methodology to jointly compute probabilistic controlled invariant  sets and state feedback controllers for nonlinear systems modeled by GPSSMs. Here, the probabilistic uncertainty is attributed to unknown components of the model as well as noise. In contrast to standard approaches where the uncertainty is independent and identically distributed (i.i.d.) noise, the uncertainty in a GPSSM is a nonlinear function of both the current state and the current input. Accordingly, the synthesized controller guarantees that the trajectory remains within the associated set with a certain probability at all times.

We first present methods for establishing probabilistic positive invariance of sets. We then propose an optimization problem for computing such sets along with state feedback controllers. Lastly, we demonstrate the applicability of our results on a physical experimental platform (quadrotor) in addition to Monte Carlo simulation.
 
  Section \ref{Gaussian Process State Space Models} introduces the system model, GPSSMs, and formulates the main problem. Section \ref{Robust and Probabilistic Invariance} investigates probabilistic invariance, providing a more general formulation and proof details that are not present in \cite{griffioen2023probabilistic,hewing2018correspondence}. Section \ref{Probabilistic Controlled Invariant Sets}  designs safety controllers for probabilistic controlled invariant sets.
Section \ref{Case Studies} validates the results on a quadrotor with simulation as well as physical experiments. Section \ref{Conclusion} gives conclusions.

\textbf{Notation}
We let $\oplus$ represent the Minkowski sum and $\|\cdot\|_2$ the Euclidean norm. Given a random variable $s$, we let $\mathbb{E}[s]$ and $\text{Cov}[s]$ represent its expected value and covariance, respectively. We let $\text{Diag}(s_1,\cdots,s_{\bar{n}})$ represent a diagonal matrix with elements $s_1,\cdots,s_{\bar{n}}$ on the diagonal. We also let $\text{BlkDiag}(S_1,\cdots,S_{\bar{n}})$ represent a block diagonal matrix with matrices $S_1,\cdots,S_{\bar{n}}$ on each block.

\section{Gaussian Process State Space Models}
\label{Gaussian Process State Space Models}

\subsection{Prior Model}
We consider the discrete-time system model that  represents uncertainty with $n$ independent GPs:
\begin{equation}
\label{Dynamics}
x_{k+1} = g(x_k,u_k) + w_k.
\end{equation}
Here $x_k\in\mathbb{R}^n$ represents the system state at time step $k$ and $u_k\in U\subseteq\mathbb{R}^m$ is the control input.  The term $w_k\sim\mathcal{N}(0,Q)$ is i.i.d. GP noise with $Q\triangleq\text{Diag}(\sigma_1^2,\cdots,\sigma_n^2)$. The term $g(x_k,u_k)$ is defined as
\begin{align}
& g(x_k,u_k) \triangleq
\begin{bmatrix}
g_1(x_k,u_k) & \cdots & g_n(x_k,u_k)
\end{bmatrix}^T,
\\
\label{GPDefinition}
&g_i(x_k,u_k) \sim \mathcal{GP}(m_i(\hat{x}_k),k_i(\hat{x}_k,\hat{x}_k')),
\end{align}
where $~\hat{x}_k\triangleq\begin{bmatrix}x_k^T&u_k^T\end{bmatrix}^T$ and $g_i(x_k,u_k)$ is a GP specified by its mean function $m_i(\hat{x}_k)$: $\mathbb{R}^{n+m}\to\mathbb{R}$ and covariance function $k_i(\hat{x}_k,\hat{x}_k')$: $\mathbb{R}^{n+m}\times\mathbb{R}^{n+m}\to\mathbb{R}$. These are given by
\begin{align}
m_i(\hat{x}_k) & \triangleq A_ix_k+B_iu_k,
\\
k_i(\hat{x}_k,\hat{x}_k') & \triangleq \mathbb{E}[(g_i(\hat{x}_k)-m_i(\hat{x}_k))(g_i(\hat{x}_k')-m_i(\hat{x}_k'))],
\end{align}
where $A_i$ and $B_i$ denote the $i^{th}$ rows of $A$ and $B$, respectively. 
We assume that $A$, $B$, $g$, and $Q$ are unknown.
A GP is a distribution over functions, assigning a joint Gaussian distribution to any finite subset of the state and control input space \cite{rasmussen2006gaussian}. The covariance function, also called the kernel, determines the class of functions over which the distribution is defined.

\subsection{Posterior Model}
We now assume that $N$ measurements of the state are taken, either through recorded trajectory data or simply by sampling the state transition function at various points in the state and control input space. This training data set, composed of $N$ data pairs, is given by $\mathcal{D}\triangleq\{\{\bar{x}_j,\bar{u}_j\},\bar{x}_j^+\}_{j=1}^N$, where
\begin{equation}
\bar{x}_j^+ = g(\bar{x}_j,\bar{u}_j) + w_j, \quad w_j \sim \mathcal{N}(0,Q),
\end{equation}
with $w_j$ being the observation noise. The training data can be used to determine the values of the hyperparameters for the mean function and the covariance function by optimizing the marginal likelihood \cite{rasmussen2006gaussian}.  

Given input training data $\{\bar{x}_j,\bar{u}_j\}_{j=1}^N$ and output training data $\{\bar{x}_j^+\}_{j=1}^N$, $g(x_k,u_k)$ conditioned on $x_k$, $u_k$, and $\mathcal{D}$ follows a Gaussian distribution, given by
\begin{equation}
\label{ConditionalDistribution}
g(x_k,u_k)|\{x_k,u_k,\mathcal{D}\}\sim\mathcal{N}(\mu(\hat{x}_k),\Sigma(\hat{x}_k)),
\end{equation}
\begin{equation*}
\mu(\hat{x}_k) \triangleq
\begin{bmatrix}
m_1(\hat{x}_k) + \bar{k}_1(\hat{x}_k)^T(K_1+\sigma_1^2I_N)^{-1}(y_1-\bar{y}_1) \\
\vdots \\
m_n(\hat{x}_k) + \bar{k}_n(\hat{x}_k)^T(K_n+\sigma_n^2I_N)^{-1}(y_n-\bar{y}_n)
\end{bmatrix},
\end{equation*}
\begin{equation*}
\Sigma(\hat{x}_k) \triangleq \text{Diag}(\xi_1(\hat{x}_k),\cdots,\xi_n(\hat{x}_k)),
\end{equation*}
\begin{equation*}
\xi_i(\hat{x}_k) \triangleq k_i(\hat{x}_k,\hat{x}_k) - \bar{k}_i(\hat{x}_k)^T(K_i+\sigma_i^2I_N)^{-1}\bar{k}_i(\hat{x}_k),
\end{equation*}
\begin{equation*}
K_i \triangleq
\begin{bmatrix}
k_i(\hat{\bar{x}}_1,\hat{\bar{x}}_1) & \cdots & k_i(\hat{\bar{x}}_1,\hat{\bar{x}}_N) \\
\vdots & \ddots & \vdots \\
k_i(\hat{\bar{x}}_N,\hat{\bar{x}}_1) & \cdots & k_i(\hat{\bar{x}}_N,\hat{\bar{x}}_N)
\end{bmatrix}, ~
\hat{\bar{x}}_j \triangleq
\begin{bmatrix}
\bar{x}_j \\
\bar{u}_j
\end{bmatrix},
\end{equation*}
\begin{equation*}
\bar{k}_i(\hat{x}_k) \triangleq
\begin{bmatrix}
k_i(\hat{\bar{x}}_1,\hat{x}_k) \\
\vdots \\
k_i(\hat{\bar{x}}_N,\hat{x}_k)
\end{bmatrix}, ~
y_i \triangleq
\begin{bmatrix}
\bar{x}_1^+(i) \\
\vdots \\
\bar{x}_N^+(i)
\end{bmatrix}, ~
\bar{y}_i \triangleq
\begin{bmatrix}
m_i(\hat{\bar{x}}_1) \\
\vdots \\
m_i(\hat{\bar{x}}_N)
\end{bmatrix},
\end{equation*}
where $\bar{x}_j^+(i)$ denotes the $i^{th}$ dimension of $\bar{x}_j^+$.

We then rewrite the system dynamics in \eqref{Dynamics} as
\begin{equation}
\label{OverallDynamics}
x_{k+1} = Ax_k + Bu_k + \hat{\mu}(\hat{x}_k) + \begin{bmatrix}I_n&I_n\end{bmatrix} \bar{w}_k(\hat{x}_k),
\end{equation}
where\begin{align}
& \hat{\mu}(\hat{x}_k) \triangleq\mu(\hat{x}_k)-\begin{bmatrix}m_1(\hat{x}_k)&\cdots&m_n(\hat{x}_k)\end{bmatrix}^T,\\
& \bar{w}_k(\hat{x}_k)\triangleq\begin{bmatrix}\bar{g}(x_k,u_k)^T&w_k^T\end{bmatrix}^T,\\
& \bar{g}(x_k,u_k)| \{x_k,u_k,\mathcal{D}\} \sim \mathcal{N}(0,\Sigma(\hat{x}_k)).
\end{align}
Note that \eqref{OverallDynamics} is a nonlinear system with a nonlinear uncertainty that is not i.i.d. This is due to the fact that both $\hat{\mu}(\hat{x}_k)$ and the uncertainty $\bar{w}_k(\hat{x}_k)$ are nonlinear functions of the current state $x_k$ and the current input $u_k$. 

The following assumption allows us to apply our sufficient conditions for probabilistic invariance to the model \eqref{OverallDynamics}.
\begin{assumption}
\label{StationaryAssumption}
The GP $g$  in \eqref{Dynamics} and, thus, $\bar{g}$ in \eqref{OverallDynamics} have stationary covariance functions.
\end{assumption}
Commonly used covariance functions are indeed stationary, such as exponential, squared exponential, $\gamma$-exponential, rational quadratic, M\'{a}tern, and others \cite[Table 4.1]{rasmussen2006gaussian}.

It follows from Assumption \ref{StationaryAssumption} that 
\begin{equation} \label{kappa-def}
\kappa_i\triangleq k_i(\alpha,\alpha)
\end{equation}
is a constant that does not depend on the argument $\alpha$.
 This allows us to bound $\hat{\mu}(\hat{x}_k)$
in \eqref{OverallDynamics} as follows:
\begin{lemma}
\label{ProbabilisticInvarianceLemma1}
For all
$k\geq0$,
\begin{equation}
\label{MeanRegion}
\|\hat{\mu}(\hat{x}_k)\|^2_2\le \phi \triangleq
 \sum_{i=1}^n \kappa_i (y_i-\bar{y}_i)^T
(K_i+\sigma_i^2I_N)^{-1}(y_i-\bar{y}_i).
\end{equation}
\end{lemma}
\begin{proof} The $i^{th}$ entry of  $\hat{\mu}(\hat{x}_k)$ is given by
\begin{equation}
\hat{\mu}_i(\hat{x}_k)=\bar{k}_i(\hat{x}_k)^T(K_i+\sigma_i^2I_N)^{-1}(y_i-\bar{y}_i).
\end{equation}
It follows from the Cauchy–Schwarz inequality  that
\begin{equation}
\label{entrywise-bound}
\begin{split}
|\hat{\mu}_i(\hat{x}_k)|^2 & \le \bar{k}_i(\hat{x}_k)^T(K_i+\sigma_i^2I_N)^{-1}\bar{k}_i(\hat{x}_k) \\
& \quad \cdot \, (y_i-\bar{y}_i)^T(K_i+\sigma_i^2I_N)^{-1}(y_i-\bar{y}_i) \\ 
& \le k_i(\hat{x}_k,\hat{x}_k)(y_i-\bar{y}_i)^T(K_i+\sigma_i^2I_N)^{-1}(y_i-\bar{y}_i) \\
& = \kappa_i (y_i-\bar{y}_i)^T(K_i+\sigma_i^2I_N)^{-1}(y_i-\bar{y}_i),
\end{split}
\end{equation}
where the second inequality holds since the posterior covariance matrix $\Sigma(\hat{x}_k)$ is positive semidefinite, and the final equality is due to (\ref{kappa-def}).
Summing (\ref{entrywise-bound}) over $i$ yields (\ref{MeanRegion}).
\end{proof}

\subsection{Problem Formulation}
Let the safety constraints be given by
\begin{equation}
\label{SafetyConstraints}
x_k\in X \triangleq \{x\in\mathbb{R}^n|\beta_i^Tx\leq1,~i=1,\cdots,n_x\} ~ \forall k\ge 0,
\end{equation}
where $n_x$ denotes the number of constraints and $\beta_i$ represents the $i^{th}$ state constraint. Our goal is to design a state-feedback controller $u_k=Lx_k$ that satisfies this safety constraint in a probabilistic sense while restricting the inputs by 
\begin{equation}
\label{InputConstraints}
u_k\in U \triangleq \{u\in\mathbb{R}^m|\zeta_i^Tu\leq1,~i=1,\cdots,n_u\} ~ \forall k\ge 0,
\end{equation}
where $n_u$ denotes the number of constraints and $\zeta_i$ represents the $i^{th}$ input constraint.
In addition, we wish to maximize the probability of satisfying these constraints:
\begin{problem}
\label{ControlDesignProblem}
Given the system in \eqref{OverallDynamics}, design a state-feedback controller
\(u_k = L x_k\) and compute an admissible set of initial conditions \(X_0 \subseteq X\)
such that, for all \(x_0 \in X_0\),
\begin{equation}
\text{Pr}( x_k \in X,\; u_k \in U) \ge p \quad \forall k \ge 0,
\end{equation}
and choose \(L\)  to make \(p\) as large as possible.
\end{problem}


\section{Probabilistic Positive Invariance}
\label{Robust and Probabilistic Invariance}
\subsection{Ellipsoids and Confidence Regions}
Before introducing probabilistic positive invariance, we define ellipsoids and confidence regions and state their properties, which we use later.
\begin{definition}
\label{EllipsoidDefinition}
An ellipsoid is given by
\begin{equation}
\mathcal{E}({\bar{\mu}},{\bar{\Sigma}}) \triangleq {\bar{\mu}} \oplus \left\{{\bar{\Sigma}}^{\frac{1}{2}}s: \|s\|_2\leq1\right\},
\end{equation}
where ${\bar{\mu}}$ is the center and ${\bar{\Sigma}}={\bar{\Sigma}}^T\succeq0$ is the shape matrix.
\end{definition}
When $\bar{\Sigma}$ is invertible,
this definition is equivalent to
\begin{equation}
\mathcal{E}({\bar{\mu}},{\bar{\Sigma}})=\left\{ s: (s-\bar{\mu})^T\bar{\Sigma}^{-1}(s-\bar{\mu})\le 1\right\}.
\end{equation}
The following property, proven in \cite{seeger1990direct}, plays an important role when characterizing probabilistic positively invariant sets.
\begin{lemma}[\cite{seeger1990direct}]
\label{EllipsoidAddition}
Let ${\bar{\Sigma}}_i={\bar{\Sigma}}_i^T\succeq0$, $i=1,2$. Then
\begin{equation}
\mathcal{E}(0,{\bar{\Sigma}}_1+{\bar{\Sigma}}_2) \subseteq \mathcal{E}(0,{\bar{\Sigma}}_1) \oplus \mathcal{E}(0,{\bar{\Sigma}}_2).
\end{equation}
\end{lemma}

\begin{definition}[\cite{hewing2018correspondence}]
\label{ConfidenceRegion}
Given a random variable $s\in\mathbb{R}^{n_s}$, the set $S\subset \mathbb{R}^{n_s}$ is a confidence region of probability level $p$ for random variable $s$, denoted $C_p(s)$, if
\begin{equation}
\text{Pr}(s\in S) \geq p.
\end{equation}
\end{definition}
The following lemma follows from the multidimensional Chebyshev inequality, which holds for arbitrary distributions.
\begin{lemma}
\label{PsiCorollary}
Let ${\bar{\mu}}=\mathbb{E}[s]$ and ${\bar{\Sigma}}=\text{Cov}[s]$, where $s\in\mathbb{R}^{n_s}$ is a random variable. Then for any $p\in(0,1)$,
\begin{equation}
\mathcal{E}\left({\bar{\mu}},\tfrac{n_s}{1-p}{\bar{\Sigma}}\right)\text{ is a }C_p(s).
\end{equation}
\end{lemma}

\subsection{Probabilistic Positive Invariance  and a Sufficient Condition}
Consider a system of the form
\begin{equation}
\label{SystemDynamics2}
z_{k+1} = \mathcal{A}_kz_k + \mathcal{B}_kd_k + \mathcal{C}_kv_k(z_k),
\end{equation}
where $z_k\in\mathbb{R}^{n}$, $d_k\in D_k\subset\mathbb{R}^{n_d}$ is an unknown but bounded term, and $v_k(z_k)\in\mathbb{R}^{n_v}$ is a random variable.
\begin{definition}[\cite{hewing2018correspondence}]
\label{PInvDef}
$Z\subset \mathbb{R}^n$ is a probabilistic positively invariant set with  probability $p\in(0,1]$ for \eqref{SystemDynamics2} if
\begin{equation}
\label{ProbabilisticInvarianceDefinition} \text{Pr}(z_k\in Z|z_0\in Z) \geq p \quad \forall k \geq 1.
\end{equation}
Equivalently, $Z$ is a $C_p(z_k)$ $\forall k\geq 1$ whenever $z_0\in Z$.
\end{definition}

The following theorem provides sufficient conditions ensuring probabilistic positive invariance of a set $Z$.

\begin{theorem}
\label{RIStoPISTheorem}
Consider the system in \eqref{SystemDynamics2}, and 
suppose
\begin{equation}\label{0-conditionaal-mean}
\mathbb{E}[v_k(z_k)|z_k]=0 \quad \forall k\geq0,
\end{equation}
 and there exist \(\bar{\Sigma}_k^v\succeq 0\), $k\ge 0$, such that
 \begin{equation}
    \label{conditional-covariance-bound} 
\text{Cov}[v_k(z_k)|z_k] \preceq \bar{\Sigma}_k^v \quad \forall k\geq0.
 \end{equation}
Let   $p\in (0,1)$ and define the ellipsoid
\begin{equation}\label{Vkdef}
V_k(p)\triangleq\mathcal{E}\left(0,\frac{n}{1-p}\bar{\Sigma}_k^v\right). 
\end{equation}
 Then a set $Z\subset \mathbb{R}^{n}$ satisfying
 \begin{equation}
\label{SubsetRelationship}
\mathcal{A}_kZ \oplus \mathcal{B}_kD_k \oplus \mathcal{C}_kV_k(p) \subseteq Z \quad \forall k\geq0
\end{equation}
is a probabilistic positively invariant set with probability $p$.
\end{theorem}

Before proving the theorem, we note that the main condition~(\ref{SubsetRelationship}) is equivalent to a robust positive invariance property for the deterministic system:
\begin{equation}
\label{SystemDynamics3}
\bar{\xi}_{k+1} = \mathcal{A}_k\bar{\xi}_k + \mathcal{B}_k d_k + \mathcal{C}_k \upsilon_k,
\end{equation}
which differs from~\eqref{SystemDynamics2} only in that $\upsilon_k\in V_k(p)\subset\mathbb{R}^{n_v}$ is an unknown but bounded term. This equivalence will allow us to ensure probabilistic invariance by designing a controller using robustness methods for deterministic systems.

\begin{proof}
We let $\mu_k^z\triangleq\mathbb{E}[z_k]$ and $\Sigma_k^z\triangleq\text{Cov}[z_k]$, and we define the sequence of sets
\begin{equation}\label{ellipZk}\bar{Z}_k\triangleq \mathcal{E}\left(\mu_k^z,\frac{n}{1-p}\Sigma_k^z\right) \ k\ge 0,
\end{equation}
where $\bar{Z}_0=\{z_0\}$ since $z_0$ is non-random, i.e., $\mu_0^z=z_0$ and $\Sigma_0^z=0$.
It follows from Lemma~\ref{PsiCorollary} that $\bar{Z}_k$ is a $C_p(z_k)$ $\forall k\geq 1$. In addition, we claim that
\begin{equation}
\label{SufficientRIStoPIS}
\bar{Z}_{k+1} \subseteq \mathcal{A}_k\bar{Z}_k \oplus \mathcal{B}_kD_k \oplus \mathcal{C}_kV_k(p) \quad \forall k\geq0.
\end{equation}
Taking \eqref{SufficientRIStoPIS}   to be true for now, we note from \eqref{SubsetRelationship}  that \begin{equation}
\label{Implication7}
\bar{Z}_k\subseteq Z \implies \bar{Z}_{k+1}\subseteq Z \quad \forall k\geq0.
\end{equation}
Thus, if $\bar{Z}_0=\{z_0\}\subset Z$, then $\bar{Z}_k \subset Z$ $\forall k\geq 1$. Since $\bar{Z}_k$ is a $C_p(z_k)$ $\forall k\geq 1$, this inclusion guarantees that $Z$ is a $C_p(z_k)$ $\forall k\geq 1$, proving the theorem.

It remains to prove the claim 
\eqref{SufficientRIStoPIS}. Note from \eqref{SystemDynamics2} that
\begin{equation}
\label{MeanDynamics}
\begin{split}
\mu_{k+1}^z &\in \mathcal{A}_k\mu_k^z \oplus \mathcal{B}_kD_k \oplus \mathcal{C}_k\mathbb{E}[v_k(z_k)] \\
&= \mathcal{A}_k\mu_k^z \oplus \mathcal{B}_kD_k \oplus \mathcal{C}_k\mathbb{E}[\mathbb{E}[v_k(z_k)|z_k]] \\
&= \mathcal{A}_k\mu_k^z \oplus \mathcal{B}_kD_k,
\end{split}
\end{equation}
\begin{equation}
\label{ConditionalMeanDynamics}
\begin{split}
\mathbb{E}[z_{k+1}|z_k] &= \mathcal{A}_kz_k + \mathcal{B}_kd_k + \mathcal{C}_k\mathbb{E}[v_k(z_k)|z_k] \\
&= \mathcal{A}_kz_k + \mathcal{B}_kd_k,
\end{split}
\end{equation}
\begin{equation}
\label{ConditionalCovDynamics}
\text{Cov}[z_{k+1}|z_k] = \mathcal{C}_k\text{Cov}[v_k(z_k)|z_k]\mathcal{C}_k^T,
\end{equation}
where the first equality in \eqref{MeanDynamics} follows from the law of total expectation and the second equalities in \eqref{MeanDynamics} and \eqref{ConditionalMeanDynamics} follow from $\mathbb{E}[v_k(z_k)|z_k]=0$. Also note that
\begin{equation}
\label{CovarianceDynamics}
\begin{split}
\Sigma_{k+1}^z &= \text{Cov}[\mathbb{E}[z_{k+1}|z_k]] + \mathbb{E}[\text{Cov}[z_{k+1}|z_k]] \\
&= \text{Cov}[\mathcal{A}_kz_k + \mathcal{B}_kd_k] + \mathbb{E}[\mathcal{C}_k\text{Cov}[v_k(z_k)|z_k]\mathcal{C}_k^T] \\
&= \mathcal{A}_k\Sigma_k^z\mathcal{A}_k^T + \mathcal{C}_k\mathbb{E}[\text{Cov}[v_k(z_k)|z_k]]\mathcal{C}_k^T \\
&\preceq \mathcal{A}_k\Sigma_k^z\mathcal{A}_k^T + \mathcal{C}_k\bar{\Sigma}_k^v\mathcal{C}_k^T,
\end{split}
\end{equation}
where the first equality follows from the law of total covariance, the second equality follows from \eqref{ConditionalMeanDynamics} and \eqref{ConditionalCovDynamics}, the third equality follows since $d_k$ is non-random, and the inequality follows from (\ref{conditional-covariance-bound}).
Then $\forall k\geq0$,
\begin{equation}
\begin{split}
&\bar{Z}_{k+1}=\mathcal{E}\left(\mu_{k+1}^z,\tfrac{n}{1-p}\Sigma_{k+1}^z\right) = \mu_{k+1}^z \oplus \mathcal{E}\left(0,\tfrac{n}{1-p}\Sigma_{k+1}^z\right) \\
&\subseteq  \mathcal{A}_k\mu_k^z \oplus \mathcal{B}_kD_k \oplus \mathcal{E}\left(0,\tfrac{n}{1-p}(\mathcal{A}_k\Sigma_k^z\mathcal{A}_k^T+\mathcal{C}_k\bar{\Sigma}_k^v\mathcal{C}_k^T)\right) \\
&\subseteq  \mathcal{A}_k\mu_k^z \oplus \mathcal{B}_kD_k \oplus \mathcal{E}\left(0,\tfrac{n}{1-p}\mathcal{A}_k\Sigma_k^z\mathcal{A}_k^T\right) \\ & \qquad \qquad \qquad \quad \ \oplus \mathcal{E}\left(0,\tfrac{n}{1-p}\mathcal{C}_k\bar{\Sigma}_k^v\mathcal{C}_k^T\right) \\
&= \mathcal{A}_k\mathcal{E}\left(\mu_k^z,\tfrac{n}{1-p}\Sigma_k^z\right) \oplus \mathcal{B}_kD_k \oplus \mathcal{C}_k\mathcal{E}\left(0,\tfrac{n}{1-p}\bar{\Sigma}_k^v\right)\\
&=\mathcal{A}_k\bar{Z}_k \oplus \mathcal{B}_kD_k \oplus \mathcal{C}_kV_k(p),
\end{split}
\end{equation}
where the first and the penultimate equalities follow from Definition \ref{EllipsoidDefinition}, the first inequality follows from \eqref{MeanDynamics} and \eqref{CovarianceDynamics}, and the second inequality follows from Lemma~ \ref{EllipsoidAddition}. 
This proves the claim \eqref{SufficientRIStoPIS}  and, hence, the theorem.
\end{proof}

\subsection{Computation with Semidefinite Programming}

We next explore the special case of \eqref{SystemDynamics2} where the matrices are time-invariant and the unknown term $d_k$ is restricted to an ellipsoid. Under these conditions, the following theorem formulates a linear matrix inequality to identify an ellipsoidal set $Z$ satisfying the  condition (\ref{SubsetRelationship}) of Theorem \ref{RIStoPISTheorem}.
\begin{theorem}
\label{ProbabilisticInvariance}
Consider the system in \eqref{SystemDynamics2} with time-invariant matrices $\mathcal{A}_k=\mathcal{A}$, $\mathcal{B}_k=\mathcal{B}$, $\mathcal{C}_k=\mathcal{C}$, and assume (\ref{0-conditionaal-mean}) holds. 
Assume, further, that
(\ref{conditional-covariance-bound}) holds with a uniform bound
\begin{equation}\label{sigma-bound-1}
\bar{\Sigma}_k^v\preceq \Sigma_v \quad \forall k\ge 0, \quad \Sigma_v=\Sigma_v^T\succ 0,
\end{equation}
and the bound $D_k$ for the uncertain term $d_k$ satisfies \begin{equation}\label{sigma-bound-2}
D_k\subseteq \mathcal{E}(0,\Sigma_d) \quad \forall k\ge 0, \quad \Sigma_d=\Sigma_d^T\succ 0.
\end{equation}
Under these conditions, if there exist $\eta\in(0,1)$ and ${S}={S}^T\succ 0$ such that  
\begin{align}\label{LMI1}
& S-\frac{1}{\eta}\mathcal{A}S\mathcal{A}^T \succeq 0, \\
& S -\frac{2}{(1-\sqrt{\eta})^2}
\left(\mathcal{B}\Sigma_d \mathcal{B}^T+
\frac{n}{1-p}\mathcal{C}\Sigma_v\mathcal{C}^T\right) \succeq 0, \label{LMI2}
\end{align}
then $\mathcal{E}(0,S)$ is  a probabilistic positively invariant set with probability $p$.
\end{theorem}

\begin{proof}
We will show that the auxiliary deterministic system in (\ref{SystemDynamics3}) satisfies the implication:
\begin{equation}\label{RIS4aux}
{\renewcommand{\arraystretch}{1.2}\begin{Bmatrix}
\bar{\xi}_k^T{S}^{-1}\bar{\xi}_k \le 1\\
d_k^T\Sigma_d^{-1} d_k\leq1\\
\frac{1-p}{n}\upsilon_k^T\Sigma_v^{-1}\upsilon_k\leq1
\end{Bmatrix}} \ \implies  \ \bar{\xi}_{k+1}^T{S}^{-1}\bar{\xi}_{k+1} \leq 1.
\end{equation}
This is equivalent to the positive invariance of  $Z=\mathcal{E}(0,{S})$ when $d_k\in \mathcal{E}(0,\Sigma_d)$ and $\upsilon_k\in \mathcal{E}\left(0,\frac{n}{1-p}\Sigma_v\right)$; that is,
\begin{equation}
\mathcal{A}Z \oplus \mathcal{B}\mathcal{E}(0,\Sigma_d) \oplus \mathcal{C}\mathcal{E}\left(0,\frac{n}{1-p}\Sigma_v\right) \subseteq Z.
\end{equation}
It then follows from (\ref{Vkdef}), (\ref{sigma-bound-1}), and (\ref{sigma-bound-2}) that condition (\ref{SubsetRelationship}) of Theorem \ref{RIStoPISTheorem} holds and, thus, $Z$ is a probabilistic positively invariant set with probability $p$ for the stochastic system \eqref{SystemDynamics2}.

We next show that (\ref{RIS4aux}) follows from
(\ref{LMI1})-(\ref{LMI2}), which are equivalent to 
\begin{align}\label{LMI3}
& \mathcal{A}^TS^{-1}\mathcal{A} \preceq \eta S^{-1} \\
& \begin{bmatrix}\mathcal{B}^T\\ \mathcal{C}^T
\end{bmatrix}
S^{-1}
\begin{bmatrix}\mathcal{B}& \mathcal{C}
\end{bmatrix} \preceq \frac{(1-\sqrt{\eta})^2}{2}
\begin{bmatrix}\Sigma_d^{-1} & 0 \\ 0 & \frac{1-p}{n}\Sigma_\upsilon^{-1}
\end{bmatrix}\label{LMI4}
\end{align}
by a Schur complement argument. Define the weighted norm
$\|\bar{\xi}\|_{{S}^{-1}}\triangleq\left( \bar{\xi}^TS^{-1}\bar{\xi}\right)^{1/2}$. Then, by the triangle inequality,
\begin{align}\nonumber
\|\bar{\xi}_{k+1}\|_{{S}^{-1}}
&=\|\mathcal{A}\bar{\xi}_k + \mathcal{B} d_k + \mathcal{C} \upsilon_k
\|_{{S}^{-1}}
\\
& \le \|\mathcal{A}\bar{\xi}_k\|_{{S}^{-1}}
+\|\mathcal{B} d_k + \mathcal{C} \upsilon_k
\|_{{S}^{-1}}.\label{triangelineq}
\end{align}
It follows from (\ref{LMI3}) that 
\begin{equation} \|\mathcal{A}\bar{\xi}_k\|_{{S}^{-1}}^2\le {\eta} \|\bar{\xi}_k\|_{{S}^{-1}}^2
\end{equation}
and from (\ref{LMI4}) that
\begin{equation}
\|\mathcal{B} d_k + \mathcal{C} \upsilon_k\|_{{S}^{-1}}^2 \le \frac{(1-\sqrt{\eta})^2}{{2}}\left(d_k^T\Sigma_d^{-1}d_k+\frac{1-p}{n}\upsilon_k^T\Sigma_\upsilon^{-1}\upsilon_k\right).
\end{equation}
Thus, when the inequalities on the left-hand side of \eqref{RIS4aux} hold, $\|\mathcal{A}\bar{\xi}_k\|_{{S}^{-1}}\le \sqrt{\eta}$
and \(\|\mathcal{B} d_k + \mathcal{C} \upsilon_k\|_{{S}^{-1}}\le 1-\sqrt{\eta}\). Then, (\ref{triangelineq}) guarantees $\|\bar{\xi}_{k+1}\|_{{S}^{-1}}\le 1$, proving
(\ref{RIS4aux}).
\end{proof}

\section{Safety Controller  Design}
\label{Probabilistic Controlled Invariant Sets}

Note that the GPSSM in (\ref{OverallDynamics}) with control $u_k=Lx_k$ is of the form \eqref{SystemDynamics2} with
\begin{equation}\label{map1}
\mathcal{A}_k=\mathcal{A}=A+BL, \ \mathcal{B}_k=\mathcal{B}=I_n, \ \mathcal{C}_k=\mathcal{C}=[I_n \ I_n].
\end{equation}
In particular, we treat $\hat{\mu}(\hat{x}_k)$ in (\ref{OverallDynamics})   as the bounded disturbance term $d_k$ in \eqref{SystemDynamics2} and rewrite the bound in Lemma~\ref{ProbabilisticInvarianceLemma1} as
\begin{equation}\label{map2}
d_k\in \mathcal{E}(0,\Sigma_d) \quad \forall k\ge 0, \quad \Sigma_d=\phi I_n.
\end{equation}
Likewise, mapping
$\bar{w}_k(\hat{x}_k)$
in (\ref{OverallDynamics})
to $v_k(z_k)$ in \eqref{SystemDynamics2}, we have
\begin{equation}
\label{map3}
\begin{split}
\text{Cov}[v_k(z_k)|z_k] &=
  \text{BlkDiag}(\Sigma(\hat{x}_k),Q)  \\
  &\preceq \text{BlkDiag}(\bar{K},Q) \triangleq  \Sigma_v,
\end{split}
\end{equation}
where $\bar{K}\triangleq \text{Diag}(\kappa_1,\cdots,\kappa_n)$.

Thus, given a controller 
$u_k=Lx_k$, we can apply Theorem~\ref{ProbabilisticInvariance} with the matrices
in (\ref{map1})-(\ref{map3}) to identify a probabilistic positively invariant set of the form $\mathcal{E}(0,S)$.
The following theorem summarizes this observation and checks the satisfaction of the state and input constraints \eqref{SafetyConstraints}, \eqref{InputConstraints}.
\begin{theorem}
\label{OverallProbabilisticInvarianceTheorem2}
Consider the system in \eqref{OverallDynamics} with control $u_k=Lx_k$ and state and input constraint sets $X$ and $U$ defined in \eqref{SafetyConstraints} and \eqref{InputConstraints}.
If there exist $\eta\in(0,1)$ and ${S}={S}^T\succ 0$ satisfying (\ref{LMI1})-(\ref{LMI2}) with the matrices
in (\ref{map1})-(\ref{map3}), and
\begin{align}
\beta_i^T{S}\beta_i &\leq 1, ~ i=1,\cdots,n_x, \label{StateConstraints} \\
\zeta_i^TLSL^T\zeta_i &\leq 1, ~ i=1,\cdots,n_u, \label{ControlConstraints}
\end{align}
then $\forall x_0\in \mathcal{E}(0,S)$,
\begin{equation}
\text{Pr}(x_k\in X,~u_k\in U)\geq p \quad \forall k\geq0.
\end{equation}
\end{theorem}
\begin{proof}
By Theorem~\ref{ProbabilisticInvariance},  $\mathcal{E}(0,S)$ is probabilistic positively invariant with probability $p$. Thus, we only need to prove $\mathcal{E}(0,S)\subseteq X$ and $L\mathcal{E}(0,S)\subseteq U$.
To prove the former we recall (see, e.g., \cite{Rockafellar1970}) that
the support function of the ellipsoid $\mathcal{E}(0,S)$ is
\begin{equation}
h_{S}(r)
\triangleq \sup_{x \in \mathcal{E}(0,S)} r^\top x=
\sqrt{r^\top  S\, r}.
\end{equation}
Thus, when $x_k\in \mathcal{E}(0,S)$,
\eqref{StateConstraints}
guarantees $\beta_i^Tx_k\le 1$, $i=1,\cdots,n_x$, ensuring $x_k\in  X$. 
Likewise, \eqref{ControlConstraints} guarantees $\zeta_i^TLx_k\le 1$, $i=1,\cdots,n_u$, which means $u_k=Lx_k\in U$.
\end{proof}

Above we assumed 
the control $u_k=Lx_k$ is given and searched for $S\succ 0$ satisfying (\ref{LMI1})-(\ref{LMI2}) and (\ref{StateConstraints})-(\ref{ControlConstraints})
with $\mathcal{A}=A+BL$
. Now we turn to the design problem where we search for $S$ and $L$ simultaneously. To do so with linear matrix inequalities, we introduce the change of variables
\begin{equation}
M\triangleq LS
\end{equation}
and use a Schur complement argument to rewrite (\ref{LMI1})
and (\ref{ControlConstraints}) as
\begin{equation}\label{LMIrewritten}
\setlength{\arraycolsep}{1pt}
    \begin{bmatrix}
S & (AS+BM)^T \\
AS+BM & \eta S
\end{bmatrix} \succeq 0, \quad \begin{bmatrix}
S & M^T\zeta_i \\
\zeta_i^TM & 1
\end{bmatrix} \succeq 0, 
\end{equation}
$i=1,\cdots,n_u$. Next, substituting  $\mathcal{B}$, $\mathcal{C}$, $\Sigma_d$, and $\Sigma_\upsilon$ from  (\ref{map1})-(\ref{map3}), we rewrite (\ref{LMI2}) as 
\begin{equation}\label{LMI2rewritten}
    S -\frac{2}{(1-\sqrt{\eta})^2}
\left(\phi I_n+
\frac{n}{1-p}(\bar{K}+Q)\right) \succeq 0 .
\end{equation}

The goal is then to find $S \succ 0$ and $M$ such that \eqref{StateConstraints}, \eqref{LMIrewritten}, and \eqref{LMI2rewritten} hold and to
recover $L$ from $L=MS^{-1}$.
The following theorem summarizes this procedure while maximizing $p$, thus offering a solution to Problem 1.

\begin{theorem}
\label{PCIDesign}
Consider the system in \eqref{OverallDynamics} where 
$\|\hat{\mu}(\hat{x}_k)\|^2_2\le \phi$ as shown in Lemma~\ref{ProbabilisticInvarianceLemma1}, 
and the state and input constraint sets $X$ and $U$ are  as defined in \eqref{SafetyConstraints} and \eqref{InputConstraints}.  Let 
$\bar{K}=\text{Diag}(\kappa_1,\cdots,\kappa_n)$ where $\kappa_i$ is as defined in (\ref{kappa-def}).
Assume the optimization problem
\begin{equation}
\label{OptimizationProblem}
\begin{split}
& \max_{p,\eta\in(0,1),S\succ 0,M} p \\
&\text{s.t.} \
\begin{bmatrix}
S & (AS+BM)^T \\
AS+BM & \eta S
\end{bmatrix} \succeq 0, \\
&\qquad S -\frac{2}{(1-\sqrt{\eta})^2}
\left(\phi I_n+
\frac{n}{1-p}(\bar{K}+Q)\right) \succeq 0, \\
&\qquad 1-\beta_i^TS\beta_i \ge  0, ~ i=1,\cdots,n_x, \\
&\qquad \begin{bmatrix}
S & M^T\zeta_i \\
\zeta_i^TM & 1
\end{bmatrix} \succeq 0, ~ i=1,\cdots,n_u,
\end{split}
\end{equation}
is feasible, and let \((p^\star,\eta^\star,S^\star,M^\star)\) be a solution. Then the controller $u_k=Lx_k$, where $L=M^\star{S^\star}^{-1}$, guarantees  that
$\forall x_0\in \mathcal{E}(0,S^\star)$, 
\begin{equation}
Pr( x_k \in X,\; u_k \in U ) \ge p^\star \quad \forall k \ge 0.
\end{equation}
\end{theorem}
Note that \eqref{OptimizationProblem} is a semidefinite program for any fixed $\eta$ and $p$ \cite{boyd1994linear}. Thus, $p$ can be maximized by applying bisection over $p$ for a set of samples $\eta\in(0,1)$. This procedure is described in Algorithm \ref{alg:opt}, where
 Line 5 can be solved in parallel for different samples of $\eta$. In addition, we maximize the size of the set $\mathcal{E}(0,S)$, which is proportional to $\log\det S$ \cite{boyd1994linear}.

\begin{algorithm}[h!]
\small
\caption{Optimization Problem Implementation}
\begin{algorithmic}[1]
\NoDo
\NoThen
\State Initialize $p$ and precision $\delta$ for the bisection over $p$
\State $p_{\text{low}}=0$, $p_{\text{up}}=1$
\While{$p_{\text{up}}-p_{\text{low}}>\delta$}	
	\For{Samples $\eta\in(0,1)$}
		\State Given $p$ and $\eta$, solve: $\argmax_{S\succ0,M}\log\det S$
		\Statex \hspace{1cm}such that the constraints in \eqref{OptimizationProblem} are satisfied
	\EndFor
	\If{Feasible for any sample $\eta$}
		\State $p_{\text{low}}=p$
	\Else
		\State $p_{\text{up}}=p$
	\EndIf
	\State $p=(p_{\text{up}}-p_{\text{low}})/2$
\EndWhile
\end{algorithmic}
\label{alg:opt}
\end{algorithm}


\section{Validation on a Quadrotor}
\label{Case Studies}
We first test the results on a high fidelity \texttt{SIMULINK} model for a quadrotor and demonstrate the probabilistic guarantee through Monte Carlo analysis. Next we present physical experiments on a quadrotor platform. 

We consider controlling a quadrotor moving in a 2-dimensional plane ($\mathsf{x}$-$\mathsf{y}$ plane) 
and model the quadrotor dynamics as a GPSSM according to \eqref{Dynamics} with states and inputs described by \cite{Ghaffari2021Analytical}. The system states and control inputs are given by \scalebox{0.9}{$x\triangleq\begin{bmatrix}x_{\mathsf{x}}&v_{\mathsf{x}}&x_{\mathsf{y}}&v_{\mathsf{y}}\end{bmatrix}^T$} and \scalebox{0.9}{$u\triangleq\begin{bmatrix}u_{\mathsf{x}}&u_{\mathsf{y}}\end{bmatrix}^T$}, where $x_i$, $v_i$, and $u_i$ are the position, velocity, and acceleration of the quadrotor on the $i^{th}$ axis, respectively, with $i\in\{\mathsf{x},\mathsf{y}\}$.

\subsection{High Fidelity SIMULINK Model}
\begin{figure}[h!]
	\centering
	\includegraphics[width=\columnwidth]{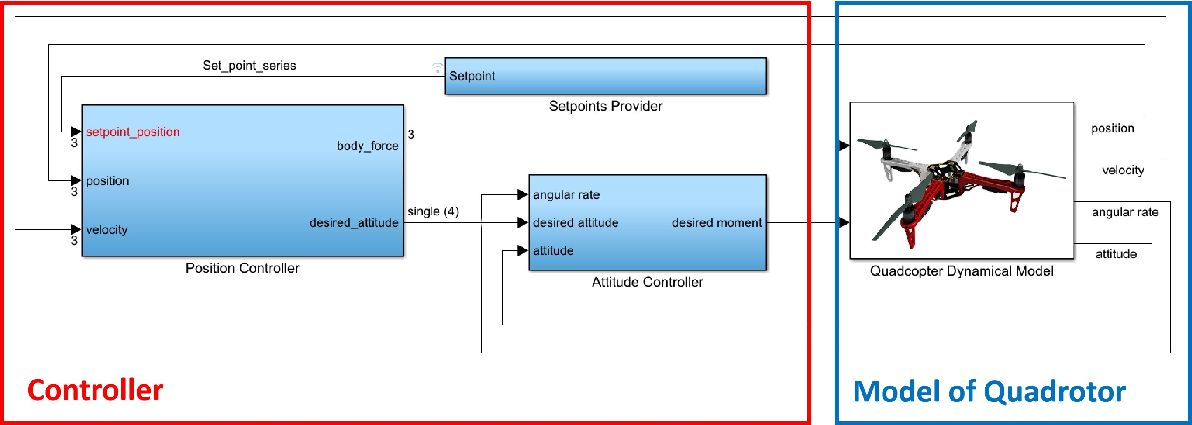}
	\caption{High fidelity \texttt{SIMULINK} for the quadrotor.}	\label{fig.controller}
\end{figure}

First, we use a high fidelity \texttt{SIMULINK} model as shown in Figure \ref{fig.controller} to validate the probabilistic guarantee proposed in this paper via Monte Carlo analysis. We consider state constraints $x_{\mathsf{x}},x_{\mathsf{y}}\in [-5,5]$ m and $v_{\mathsf{x}},v_{\mathsf{y}}\in [-7,7]$ m/s, and input constraints $u_{\mathsf{x}},u_{\mathsf{y}}\in[-5,5]$ m/s$^2$.

We collected a single state-input trajectory with $N=550$ and trained the GPSSM with a squared exponential covariance function given by
\begin{equation*}
k_i(\hat{x}_k,\hat{x}_k')=\sigma_ie^{(\hat{x}_k-\hat{x}_k')^T\Lambda_i^{-2}(\hat{x}_k-\hat{x}_k')},~i\in\{1,2,3,4\},
\end{equation*}
by using the function \texttt{fitrgp} in \texttt{MATLAB}. After this process, the trained hyperparameters are given by
\begin{align*}
A &= \begin{bmatrix}
0.9999 & 0.1009 & -0.0001 & -0.0005 \\
-0.0018 & 1.0160 & -0.0025 & -0.0086 \\
0 & 0.0008 & 0.9999 & 0.0996 \\
-0.0014 & 0.0149 & -0.0024 & 0.9926
\end{bmatrix}, \\
B &= \begin{bmatrix}
0.0028 & 0.0603 & -0.0017 & -0.0309 \\
-0.0017 & -0.0291 & 0.0028 & 0.0619
\end{bmatrix}^T,
\end{align*}
$Q=10^{-4}\text{Diag}(2.6429,2.5738,2.3335,2.5739)$, $\sigma_1=1.3343\times10^{-5}$, $\Lambda_1=9.6392\times10^{3}I_6$, $\sigma_2=1.2362\times10^{-5}$, $\Lambda_2=2.2333\times10^{3}I_6$, $\sigma_3=1.2539\times10^{-5}$, $\Lambda_3=9.6410\times10^{3}I_6$, $\sigma_4=1.1428\times10^{-5}$, and $\Lambda_4=5.0345\times10^{3}I_6$. With this model, we obtain using \texttt{Mosek} \cite{MOSEKApS2019MOSEK} and \texttt{YALMIP} the positively invariant set $\mathcal{E}(0,S)$ with $\eta=0.9251$, probability $p=0.9997$,
\begin{align*}
S &= \begin{bmatrix}
23.6862 &  -9.2063   & 1.9656 &   0.7041
 \\  -9.2063   & 9.5264 &  -2.2176  & -1.4113 \\ 
    1.9656    & -2.2176   & 24.9786   & -9.7507 \\
    0.7041   &
    -1.4113   & -9.7507   & 13.2479
\end{bmatrix}, \\
L &= \begin{bmatrix}
-0.6162 & -1.9897 & -0.3997 & -0.8999 \\
0.0297 & -0.8025 & -0.9550 & -1.5374
\end{bmatrix}.
\end{align*}

\begin{figure*}[h!]
	\centering
	\subfigure{\includegraphics[width=0.328\textwidth]{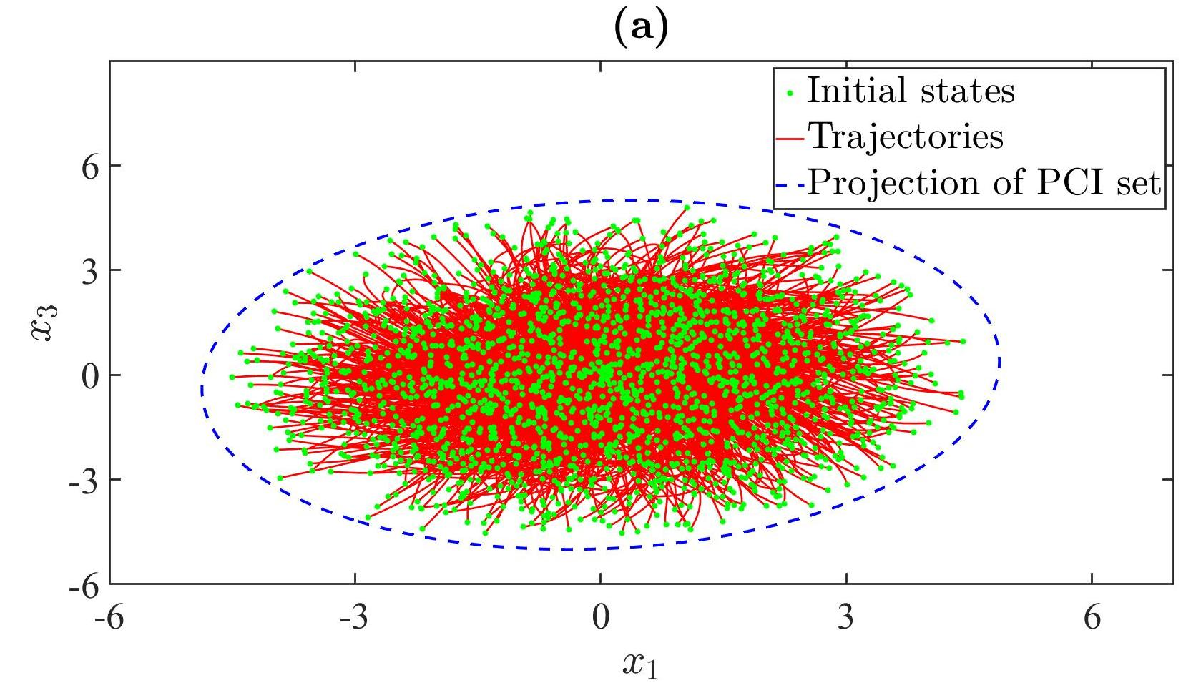}}
	\subfigure{\includegraphics[width=0.328\textwidth]{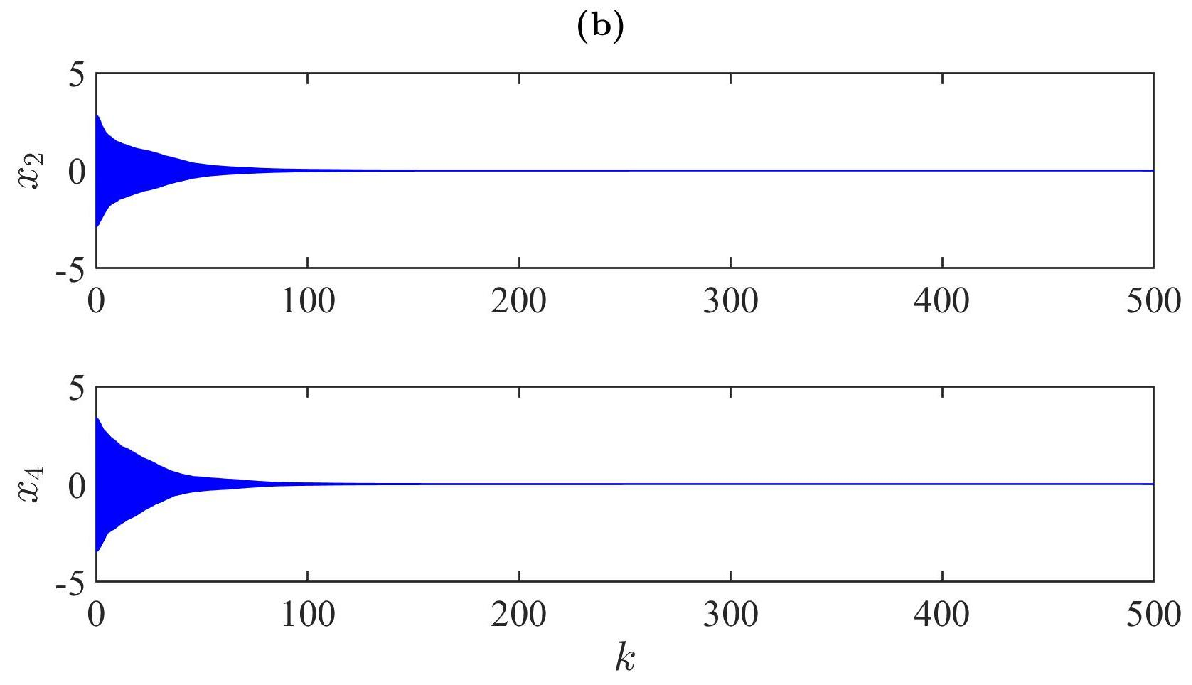}}
	\subfigure{\includegraphics[width=0.328\textwidth]{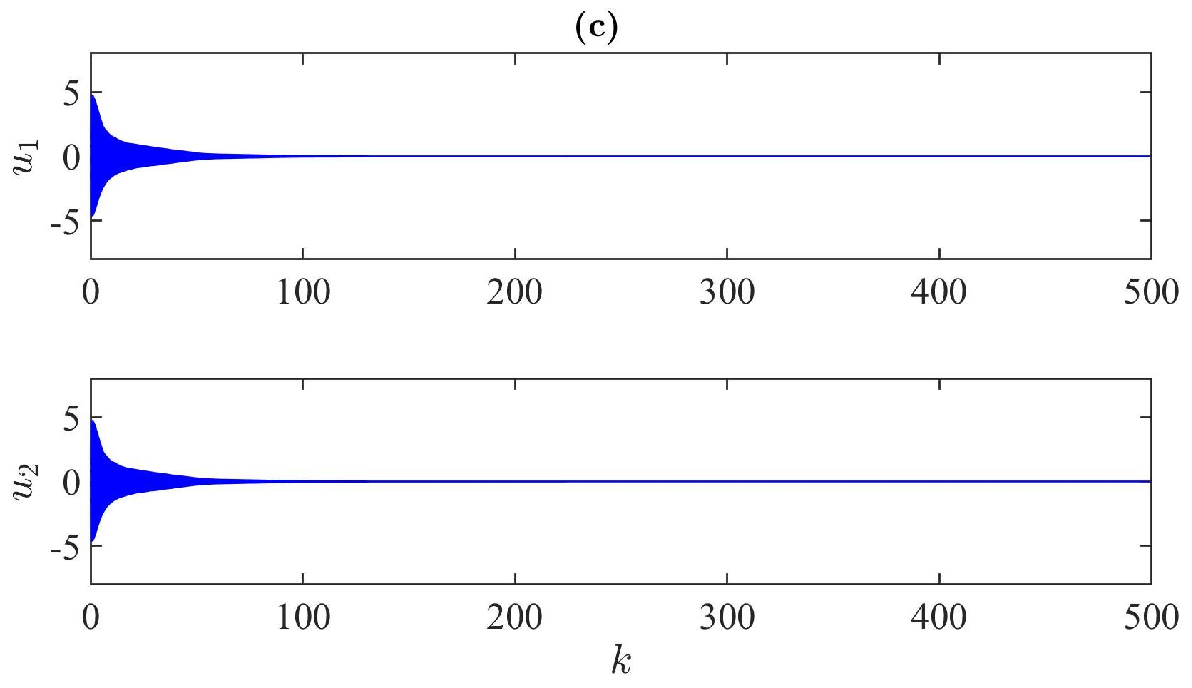}}
	\caption{Simulation results for the high fidelity \texttt{SIMULINK} model of the quadrotor case study. Figure (a) denotes the projection of the PCI set and some of the state trajectories of the quadrotor onto the $x_1$-$x_3$ plane. Figure (b) demonstrates a few sequences of the velocity of the quadrotor. Figure (c) illustrates sequences of the control inputs of the quadrotor.}
	\label{fig:monto_quad}
\end{figure*}

For the simulation, we randomly select $10^6$ initial states within the PCI set and simulate the \texttt{SIMULINK} model from each initial state for a time horizon of $T=500$.

The simulation results are given in Table \ref{tbl:compare2}, which shows that the probabilistic safety guarantees are respected.
Figure \ref{fig:monto_quad} depicts some of the state trajectories, sequences of velocities on different axes, and control input sequences.
\begin{table}[h!]
	\centering
	\renewcommand\arraystretch{1.2}
	\begin{small}
	\begin{tabular}{|c|c|}
	\hline
	$\min_{k\in[0,500]}\text{Pr}(x_k\in\mathcal{E}(0,S))$ & 100\% \\ \hline
	$\min_{k\in[0,500]}\text{Pr}(u_k\in U|x_k\in\mathcal{E}(0,S))$ & 100\% \\ \hline
	Pr($x_k\in X~\forall k\in[0,500]$) & 100\% \\ \hline
	Pr($x_k\in\mathcal{E}(0,S)~\forall k\in[0,500]$) & 99.99\% \\ \hline
	\end{tabular}
	\end{small}
	\caption{Results of Monte Carlo analysis over the high fidelity \texttt{SIMULINK} model of the quadrotor using the safety controller.}
	\label{tbl:compare2}
\end{table}

\subsection{Experimental  Testbed}
\begin{figure}[h!]
	\centering
	\subfigure{\includegraphics[width=0.49\columnwidth]{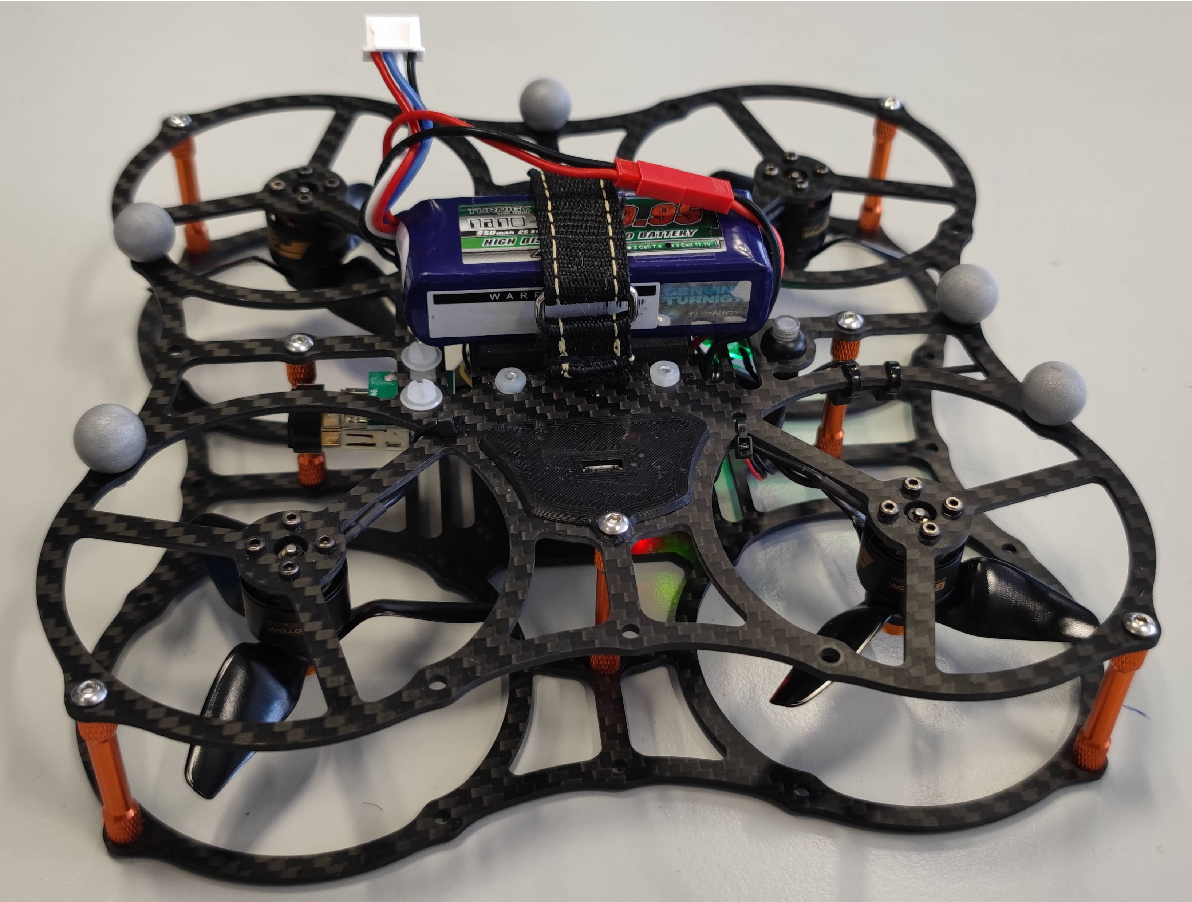}}
	\subfigure{\includegraphics[width=0.49\columnwidth]{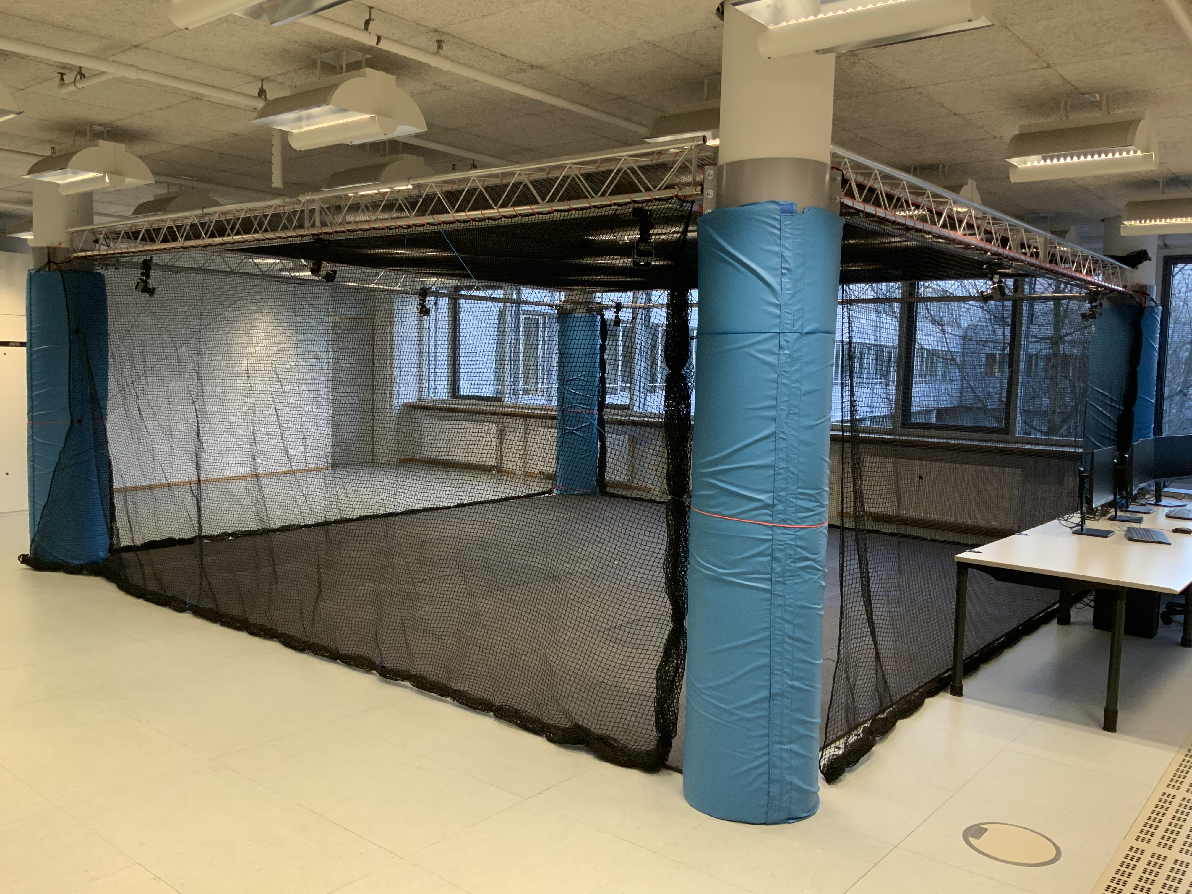}}
	\caption{Left: Physical quadrotor for the experiment. Right: Test field equipped with a motion capture system and a ground control station.}
	\label{fig:physical_platform}
\end{figure}

The physical testbed includes a quadrotor (Figure \ref{fig:physical_platform}, left) and a test field (Figure \ref{fig:physical_platform}, right)  equipped with a motion capture system for recording the position and velocity of the quadrotor and a ground control station (GCS) for running the controller and sending the desired accelerations (i.e., the control input) to the quadrotor at runtime.

According to the setting of the physical quadrotor and the spatial restrictions of the physical test field, we consider state constraints $x_{\mathsf{x}},x_{\mathsf{y}}\in[-2.5,2.5]$ m and $v_{\mathsf{x}},v_{\mathsf{y}}\in[-7,7]$ m/s, and input constraints $u_{\mathsf{x}},u_{\mathsf{y}}\in[-7,7]$ m/s$^2$.  We collect a single state-input trajectory from the quadrotor and construct a data set containing $N=436$ data points. Based on this data set, we train the GPSSM with a squared exponential covariance function given by
\begin{equation*}
k_i(\hat{x}_k,\hat{x}_k')=\sigma_ie^{(\hat{x}_k-\hat{x}_k')^T\Lambda_i^{-2}(\hat{x}_k-\hat{x}_k')},~i\in\{1,2,3,4\},
\end{equation*}
by using the function \texttt{fitrgp} in \texttt{MATLAB}. After this procedure, the trained hyperparameters are given by
\begin{align*}
A &= \begin{bmatrix}
1.0002 & 0.1009 & -0.0001 & -0.0002 \\
-0.0023 & 0.9919 & -0.0001 & 0.0013 \\
-0.0005 & 0.0003 & 1.0001 & 0.1011 \\
-0.0039 & 0.0025 & 0.0001 & 0.9901
\end{bmatrix}, \\
B &= \begin{bmatrix}
0.0046 & 0.0720 & -0.0018 & -0.0112 \\
-0.0005 & -0.0062 & 0.0041 & 0.0844
\end{bmatrix}^T,
\end{align*}
$Q=10^{-3}\text{Diag}(0.0233,0.1210,0.0815,0.1679)$, $\sigma_1=0.0014$, $\Lambda_1=8.2882\times10^{4}I_6$, $\sigma_2=0.0034$, $\Lambda_2=491.0781I_6$, $\sigma_3=7.7636\times10^{-6}$, $\Lambda_3=3.3397\times10^{7}I_6$, $\sigma_4=0.0063$, and $\Lambda_4=644.1183I_6$. By following the same synthesis procedure for the high fidelity model, we obtain $\eta=0.8230$, probability level $p=0.9736$, 
\begin{align*}
S &= \begin{bmatrix}
6.2532  & -6.0216 &   0.4840  & -0.1400 \\
   -6.0216   & 13.6011  & -0.8377  &  0.2895 \\
    0.4840  & -0.8377  &  6.2512  & -6.0153\\
   -0.1400  &  0.2895 &  -6.0153  & 15.5384
\end{bmatrix}, \\
L &= \begin{bmatrix}
-2.6987 & -2.4831 & -0.1247 & -0.1870 \\
-0.2101 & -0.4243 & -2.7680 & -2.1569
\end{bmatrix}.
\end{align*}

\begin{figure}[h!]
	\centering
	\subfigure{\includegraphics[width=\columnwidth]{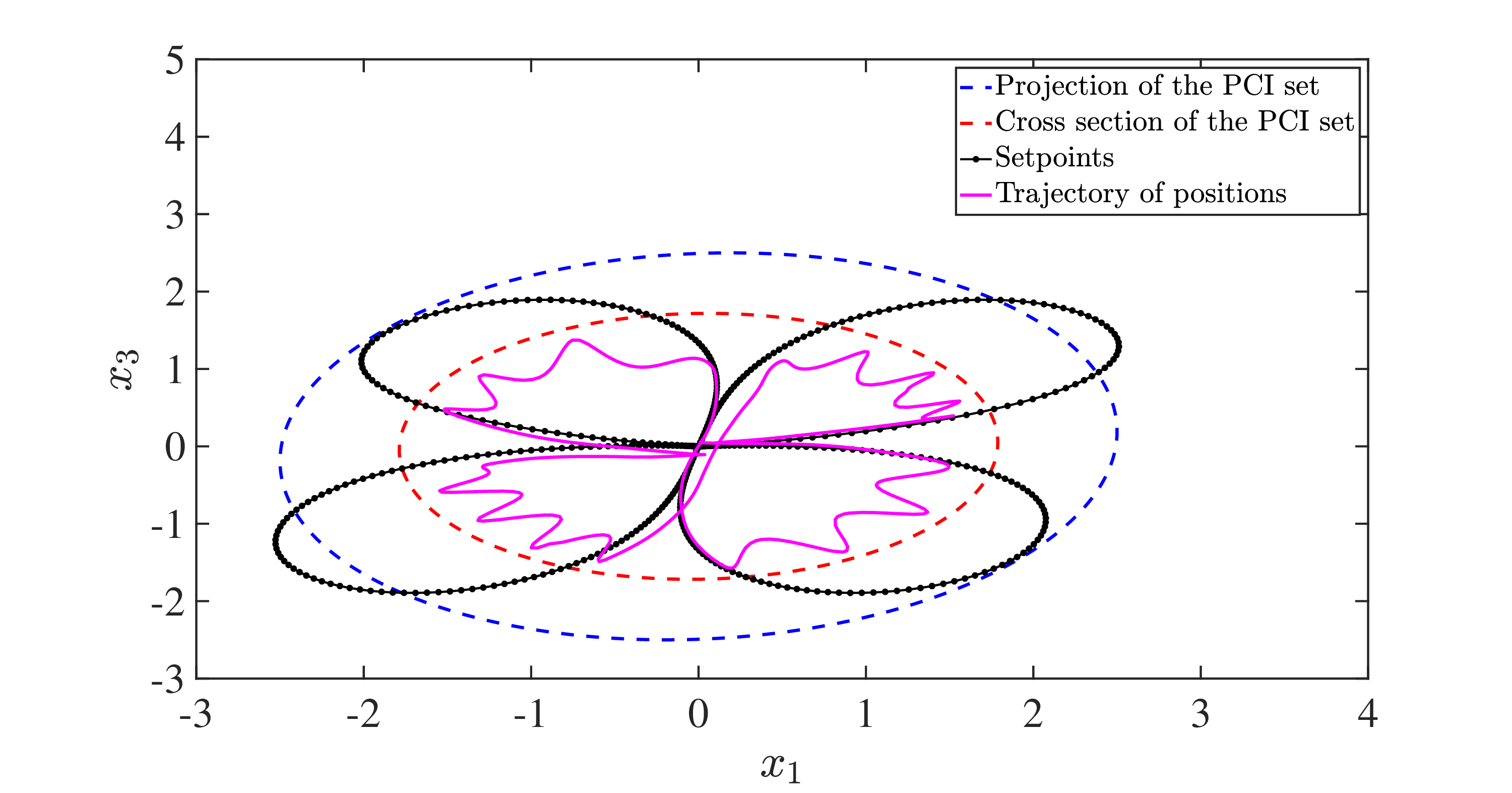}}
	\subfigure{\includegraphics[width=\columnwidth]{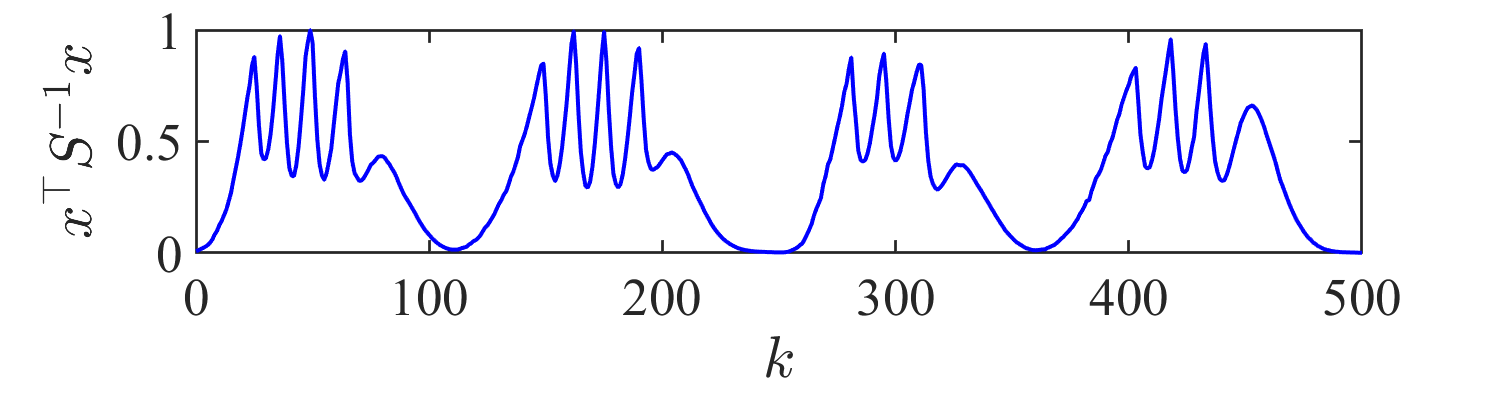}}
	\caption{Top: Evolution of the quadrotor's positions in the real-world experiment.
	Bottom: Values of $x_k^TS^{-1}x_k$ along the trajectory of the quadrotor.}
	\label{fig:physical_experiment state}
\end{figure}

\begin{figure}[h!]
	\centering
	\subfigure{\includegraphics[width=\columnwidth]{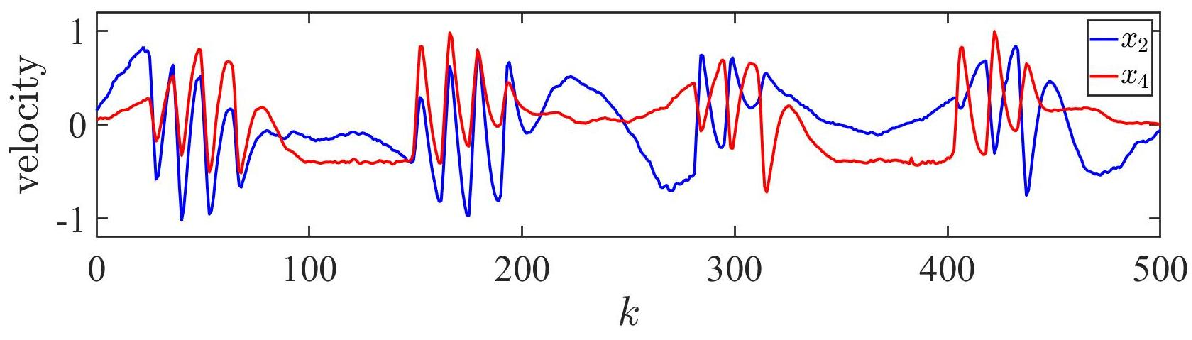}}
	\subfigure{\includegraphics[width=\columnwidth]{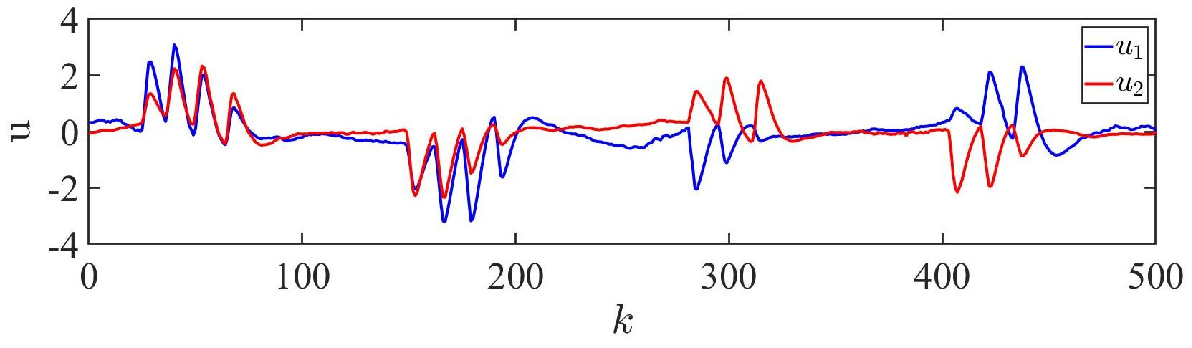}}
	\caption{Sequences of the quadrotor's velocity and control inputs in the real-world experiment.}
	\label{fig:physical_experiment input}
\end{figure}

In Figure \ref{fig:physical_experiment state}, we plot the quadrotor's trajectory over 50 s (i.e., $T=500$) as it follows a series of set points $x_{\text{set}}$ that go outside the positively invariant set $\mathcal{E}(0,S)$. This set is deployed as a safety filter, where a set-point tracking controller $u=L_{\text{set}}(x_k-x_{\text{set}})$ with
\begin{equation*}
L_{\text{set}} = \begin{bmatrix}
-1.4781 & -1.7309 & 0 & 0 \\
0 & 0 & -1.4781 & -1.7309
\end{bmatrix}
\end{equation*}
is the prime controller. The state-feedback controller $u_k=Lx_k$ associated with $\mathcal{E}(0,S)$ is then used as the backup controller that is applied when the set-point tracking controller is driving the system outside this set. In Figures \ref{fig:physical_experiment state} and \ref{fig:physical_experiment input} we depict the position sequence and the sequences of velocity and control inputs of the quadrotor, respectively. As can be seen in the figures, the desired safety and input constraints are respected while the state trajectory of the quadrotor stays within the positively invariant set. A video for the experiment is available online: \url{https://youtu.be/mEuuRIm57j4}.

\section{Conclusion}
\label{Conclusion}
We proposed an optimization-based method for synthesizing probabilistic positively invariant sets with state-feedback controllers for GPSSMs. These controllers provide safety guarantees for nonlinear systems with unmodeled and unknown dynamics. 
The results are validated on a quadrotor in both high-fidelity simulations and a physical experiment. One research direction is designing controllers 
that
provide safety guarantees while also maximizing performance.
Another direction is leveraging reachability analysis over GPSSMs to enlarge the size of positively invariant sets. Lastly,  methods for computing the covariance matrix online should be investigated.



\bibliographystyle{cas-model2-names}

\bibliography{root}

\end{document}